\documentclass[sigconf]{acmart}

\AtBeginDocument{%
  }

\copyrightyear{2024}
\acmYear{2024}
\setcopyright{acmlicensed}
\acmConference[CIKM '24]{Proceedings of the 33rd ACM International Conference on Information and Knowledge Management}{October 21--25, 2024}{Boise, ID, USA} 
\acmBooktitle{Proceedings of the 33rd ACM International Conference on Information and Knowledge Management (CIKM '24), October 21--25, 2024, Boise, ID, USA}
\acmDOI{10.1145/3627673.3680047} 
\acmISBN{979-8-4007-0436-9/24/10}

\usepackage{hyperref}       
\usepackage{url}            
\usepackage{nicefrac}       
\usepackage{microtype}      
\usepackage{algorithm}  
\usepackage{algpseudocode}
\usepackage{graphicx}
\usepackage{enumitem}
\usepackage{multirow}

\begin{document}

\title{Enhancing Relevance of Embedding-based Retrieval at Walmart}



\author{Juexin Lin}
\affiliation{%
  \institution{Walmart Global Technology}
  \city{Sunnyvale}
  \country{USA}
}
\email{juexin.lin@walmart.com}

\author{Sachin Yadav}
\affiliation{%
  \institution{Walmart Global Technology}
  \city{Bangalore}
  \country{India}
}
\email{sachin.yadav1@walmart.com}

\author{Feng Liu}
\email{f.liu@walmart.com}

\author{Nicholas Rossi}
\email{nicholas.rossi@walmart.com}
\affiliation{%
  \institution{Walmart Global Technology}
  \city{Sunnyvale}
  \country{USA}
}

\author{Praveen R. Suram}
\affiliation{%
  \institution{Walmart Global Technology}
  \city{Bangalore}
  \country{India}
}
\email{praveen.suram@walmart.com}

\author{Satya Chembolu}
\email{satya.chembolu@walmart.com}

\author{Prijith Chandran}
\email{prijith.chandran@walmart.com}

\author{Hrushikesh Mohapatra}
\email{hrushikesh.mohapatra@walmart.com}
\affiliation{%
  \institution{Walmart Global Technology}
  \city{Sunnyvale}
  \country{USA}
}

\author{Tony Lee}
\email{tony.lee@walmart.com}

\author{Alessandro Magnani}
\email{alessandro.magnani@walmart.com}

\author{Ciya Liao}
\email{ciya.liao@walmart.com}
\affiliation{%
  \institution{Walmart Global Technology}
  \city{Sunnyvale}
  \country{USA}
}

\renewcommand{\shortauthors}{Juexin Lin et al.}

\begin{abstract}
Embedding-based neural retrieval (EBR) is an effective search retrieval method in product search for tackling the vocabulary gap between customer search queries and products. The initial launch of our EBR system at Walmart yielded significant gains in relevance and add-to-cart rates \cite{magnani2022semantic}. However, despite EBR generally retrieving more relevant products for reranking, we have observed numerous instances of relevance degradation. Enhancing retrieval performance is crucial, as it directly influences product reranking and affects the customer shopping experience. Factors contributing to these degradations include false positives/negatives in the training data and the inability to handle query misspellings. To address these issues, we present several approaches to further strengthen the capabilities of our EBR model in terms of retrieval relevance. We introduce a Relevance Reward Model (RRM) based on human relevance feedback. We utilize RRM to remove noise from the training data and distill it into our EBR model through a multi-objective loss. In addition, we present the techniques to increase the performance of our EBR model, such as typo-aware training, and semi-positive generation. The effectiveness of our EBR is demonstrated through offline relevance evaluation, online AB tests, and successful deployments to live production.
\end{abstract}


\maketitle

\section{Introduction}
Optimizing search functionality for expedited and effortless product discovery is crucial to any e-commerce platform. Traditional keyword-based search, also known as lexical inverted-index retrieval \cite{ir}, has been the backbone of e-commerce search for years. While it offers advantages such as explainability, scalability, and efficiency, the lexical approach also exhibits several limitations. These shortcomings, known as vocabulary gaps, include its inability to capture the semantic relationships between terms, thereby failing to handle synonyms, as well as its over-reliance on exact keyword matches, which consequently results in difficulty in accommodating misspellings. Embedding-based retrieval (EBR) has demonstrated its ability to overcome these limitations by representing both queries and products as dense vectors using a dual-encoder architecture for learning the proximity between queries and products. The products semantically close to the query are retrieved from a fast approximate nearest neighbor (ANN) search \cite{johnson2019billion}. 

Recently, EBR has been integrated into the e-commerce search systems of many large platforms, such as Walmart, Amazon, Facebook Marketplace, and others \cite{magnani2022semantic,nigam2019semantic,muhamed2023web,he2023que2engage,jha2023unified,li2021embedding,zhang2020towards}. Most of them focus on optimizing user engagements through contrastive loss \cite{xiong2020approximate}, by differentiating the engaged products collected from search logs from random or informative in-batch negatives. Alternatively, they also aim to enhance personalization by integrating user context. 
Given that the downstream re-ranking system might not consistently rank the EBR results accurately \cite{rossi2024relevance, gao2021complement, liu2021que2search}, there is a risk that irrelevant products retrieved through EBR could be presented to customers, thereby hindering their shopping experience. As such, it becomes essential to enhance the relevance performance of EBR. 

The introduction of EBR \cite{magnani2022semantic} on Walmart.com transformed its search functionality by integrating EBR system as a complement to the traditional lexical retrieval. This integration increased the number of relevant products presented to users and therefore increased user engagement. Nevertheless, following its implementation, numerous instances of relevance degradations have been observed through the EBR retrieval. The key factors leading to these degradation are
\begin{itemize}
\item False positives within the training data derived from customer engagement.
\item False negatives in the training data generated from offline negative generation.
\item Inability to handle misspelled queries, which are common in live traffic.
\end{itemize}
In this paper, we present several approaches to address these issues for improving the relevance of EBR, building on the work of \cite{magnani2022semantic}. 

Our general idea for addressing the data-related challenges is to leverage human relevance judgment data, which evaluates the relevance of products to specific queries. Based on this data, we develop a human-feedback relevance reward model (RRM), inspired by reinforcement learning with human feedback (RLHF) \cite{ouyang2022training} in large language models. The RRM acts as a relevance proxy for each query-product pair. To attain higher precision, we employ a cross-encoder architecture \cite{dai2019deeper} for the RRM, which takes advantage of early interactions between the query and product. 

To handle the issue of false positives, we apply the RRM to revise the training label. We propose a novel multi-objective loss, supplemented with a relevance objective based on RRM, to elevate relevance even further. To address the false negatives, we present an enhanced offline negative generation algorithm that offers better relevance guardrails. During this procedure, in addition to mining negatives from the top of the retrieved set, we propose generating semi-positive samples, which are relevant to the query, from the lower positions of the retrieved set. As a solution to the third issue, we incorporate typos into queries during training to enhance the model's robustness against misspelled queries. 

To assess the performance of the proposed methods, we introduce new evaluation metrics based on human judgments and design an evaluation method on a big index products that simulates online performance. Similar to \cite{magnani2022semantic}, we deploy our candidate models in a hybrid retrieval system setting for all traffic on walmart.com. 
The efficacy of the proposed methods is demonstrated through both end-to-end relevance evaluations and live A/B tests.

In summary, the contributions of the paper are:
\begin{itemize}
    \item We introduce a relevance reward model based on human relevance feedback, and propose a novel approach to incorporate it into model training via training label revision and a new objective that optimizes for relevance in addition to engagement.
    \item We adopt a typo-injection method by augmenting queries during training.
    \item We propose several improvements during model training, including label assignment, offline hard negative generation and semi-positive generation.
\end{itemize}

\section{Related Work}
\label{section.related}
\textit{Embedding-based retrieval.} 
The two-tower Siamese architecture~\cite{bromley1993signature} for dense retrieval was first introduced by DSSM \cite{DSSM} in the context of web search. In recent years, this technique has been adopted across various e-commerce platforms, demonstrating its effectiveness in enhancing search relevance \cite{nigam2019semantic, magnani2022semantic, zhang2020towards}. Specifically, the EBR implemented on Walmart.com \cite{magnani2022semantic} represents a competitive state-of-the-art method, which we enhance in this work by introducing several innovative concepts. 
To handle the poor-quality products from ANN retrieval, Taobao \cite{zheng2023delving} applied additional filter modules using lexical matching based on query understanding components. \cite{rossi2024relevance} proposed a query-dependent relevance filtering module at pre-ranking stage for truncating the irrelevant candidates. 

\textit{Multi-objective optimization.} 
Given the complex nature of e-commerce search systems, researchers have begun to optimize for various business objectives. MOPPR \cite{zheng2022multi} proposed a model that integrates four optimization objectives, comprising relevance, exposure, click and purchase, where the relevance label is derived from a relevance estimation model \cite{yao2021learning}, trained on user click-through data. Que2Engage \cite{he2023que2engage} employed multitask learning, which consists of a contrastive loss for relevance objectives and a BCE loss for engagement objectives. 
Our use of multi-objective optimization differs from previous works by incorporating human-judged relevance. Although the term "relevance objectives" was used in prior studies, it is intrinsically tied to customer engagements, which are independent of human judgments.


\textit{Cross-encoders.} 
Cross-encoders, which execute full token-level self-attention between query and product, achieve much higher precision than dual-encoders. However, due to the high computational cost, cross-encoder models are primarily utilized for reranking \cite{nogueira2019passage, yang2019simple}. Recently, several studies have explored integrating cross-encoders at the retrieval stage. RocketQA \cite{qu2020rocketqa} used a cross-encoder to filter out false negatives during hard negative sampling. ERNIE-Search \cite{lu2022ernie} enhanced dual-encoders using a cross-encoder distillation setting. 
To leverage the effectiveness of cross-encoder, we employ it for our RRM model and subsequently distill it into a dual-encoder architecture through a multi-objective optimization for retrieval.

\textit{Hard negative mining.} Utilizing hard negative samples is an effective strategy for learning semantic retrieval models. ANCE training by \cite{xiong2020approximate} recommended to periodically update the ANN index and select top-ranked documents as negatives. A separate work~\cite{zhan2021optimizing} proposed combining hard negative samples and easy negative samples to further improve semantic retrieval model learning. In an e-commerce setting, the application of ANCE \cite{magnani2022semantic} integrates innovative strategies to identify relevant products. Our hard negative mining approach adopts \cite{magnani2022semantic} and introduces an enhanced strategy to pinpoint the relevant products. 

\textit{Typo-aware training.} 
The authors in \cite{zhuang2021dealing} proposed a typos-aware training for BERT-based retrievers that modifies the queries in training set with a 50\% probability according to different typo types, including insertion, deletion, substitution and swap. We utilize similar training strategies, with minor adjustments tailored for e-commerce applications.


\section{A new labeling framework}
\label{section.label}
Recall that in \cite{magnani2022semantic}, the label for each query and product pair was determined using a step function based on the smoothed estimation of order rate, click-through rate and impressions. 
The limitation of this labeling system is its inability to effectively distinguish between purchased products and those that have been clicked or impressed, leading to suboptimal recall performance for purchased products. Moreover, clicks are not taken into account when products have similar order rates. To address these issues, we propose a new labeling scheme that employs a weighted summation of all orders, add-to-carts, clicks, and impressions:
\begin{equation} \label{eq.eng.label}
\begin{split}
S_{ij} =& \mbox{ } w_i * \mbox{Impressions} + w_c * \mbox{Clicks} + w_a * \mbox{ATCs} + \mbox{Orders},
\end{split}
\end{equation}
where $S_{ij}$ denotes the label of product $j$ for query $i$. The weights $w_i$, $w_c$, and $w_a$ are assigned values of $0.001$, $0.01$, and $0.1$, respectively. The low weight for impressions is intentional to differentiate impressed products from negative ones. Negative products selection and label assignments are described in section~\ref{section.negative}. 

As discussed in section~\ref{section.loss}, the loss used for the EBR model is the listwise softmax loss \cite{cao2007learning}, where the label of each query and product can be treated as the assigned weight. This new labeling scheme ensures the ordered products contribute significantly to the loss. 

\section{Relevance reward model}
\label{section.rrm}
\begin{figure}
\centering
\includegraphics[width=0.35\textwidth]{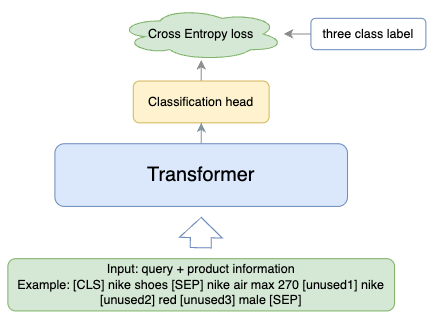}
\caption{Relevance reward model is trained with multi-class cross entropy loss leveraging the cross-encoder architecture.}
\label{fig.relmodel} 
\end{figure}

Unlike conversational chatbots, collecting human feedback for product search is not trivial. Similar to the Amazon shopping dataset \cite{reddy2022shopping}, we have a large relevance judgment dataset annotated on a 3-point scale: exact, substitute, and irrelevant. To incorporate this human feedback into our EBR, we trained a cross-encoder model for relevance judgment. First, we pre-trained a BERT base model using Walmart dataset through the self-supervised Masked Language Model task and supervised query-product classification task (highly engaged products as positive, and random products as negative). We then fine-tuned the model on human judgment data using a multi-class classification task, which included exact, substitute, and irrelevant classes. As illustrated in Figure~\ref{fig.relmodel}, this RRM takes query and product attributes designed to predict the relevance label of each prediction class. 
We incorporate the RRM into our EBR model in two different ways as detailed in section~\ref{section.score_augmentation} and section~\ref{section.rel_label}. Although both methods contribute to the improvement of relevance, the method in section~\ref{section.rel_label} turns out to be more effective than the one in section~\ref{section.score_augmentation}, as it introduces a new relevance label affecting all instances during training via loss updates (see section~\ref{result.rrm}).

\subsection{Label revision}
\label{section.score_augmentation}
Customers may sometimes purchase products not directly related to their initial search queries. This is because shopping behaviors are influenced not only by relevance, but also by factors such as product price, product image, ratings, and reviews. 
For instance, we observed purchases of a clay saucer for the query "terra cotta plant pots", possibly due to its attractive price. Another example involves customers searching for a "bird scooter for kids" but ultimately buying a refurbished adult electric scooter or an accessory handlebar. We observed a significant number of false positives within the training data, where multiple orders were made for unrelated products. To eliminate these noises, we utilized the RRM to reduce false positives by adjusting labels accordingly. The label revision scheme employs a two-stage downgrading process. We use the prediction of exact match class, denoted by $\text{P}^E_{ij}$, to downgrade the product labels to $a$ and $b$ for moderately relevant products ($0.3 \le \text{P}^E_{ij} < 0.7$) and less relevant products ($P^E_{ij} < 0.3$) respectively, with $a > b$. The thresholds of $0.3$ and $0.7$ used here are the 90th and 95th percentile values for the irrelevant class. Formally, 
\begin{equation} \label{eq.label_revision}
S^\prime_{ij} =   \left\{ \begin{array}{cl}
a & \ \mbox{if } 0.3 \le \text{P}^E_{ij} < 0.7, \mbox{ and } S_{ij} > a \\
b &  \ \mbox{if } P^E_{ij} < 0.3, \mbox{ and } S_{ij} > b \\
S_{ij} & \mbox{otherwise},
\end{array} \right.  
\end{equation}
where, $S_{ij}$ and $S^\prime_{ij}$ represent the original label and revised label of product $j$ and query $i$, respectively. Specifically, we set $a = 0.1$ (equivalent to one ATC) and $b = 0.01$ (equivalent to one click). By applying this method to our training data (section~\ref{section.data}), 3\% and 30\% of query-products pairs are downgraded to $0.1$ and $0.01$, respectively. This reveals that there is a significant amount of noisy data in our customer engagement log.  

\subsection{Relevance label}
\label{section.rel_label}
In \cite{magnani2022semantic}, the label of each query and product pair was determined by the historical customer engagements. It overly emphasized the engagement. To make the model preserve the semantic relevance besides engagement, we introduce a relevance label based on the RRM. We aim to retrieve exact matched products over substitute products and substitute products over irrelevant ones. Since RRM has three classes, we have to combine the predicted probabilities of these three classes into a single value for relevance label. In formulating the relevance label, we introduce a penalty multiplier $0 < \lambda_2 < 1$ that is applied to the probability of the substitute class if the query-product pair belongs to either the exact match or substitute class. When the pair belongs to the irrelevant class, we impose a greater penalty through another multiplier $0 < \lambda_1 < 1$. We define the relevance label for product $j$ and query $i$, denoted by $R_{ij}$, as follows.
\begin{equation} \label{eq.rel.label}
R_{ij} =  \left\{ \begin{array}{cl}
\lambda_1 * (\mbox{P}^{E}_{ij} + \lambda_2 * \mbox{P}^{S}_{ij}) & \ \mbox{if } \mbox{P}^{I}_{ij} > \max(\mbox{P}^{E}_{ij}, \mbox{P}^{S}_{ij}) \\
\mbox{P}^{E}_{ij} + \lambda_2 * \mbox{P}^{S}_{ij} &  \ \mbox{otherwise}  \
\end{array} \right.  
\end{equation}
where the $0<\lambda_1, \lambda_2 < 1$ are empirically chosen parameters, while $\mbox{P}^{E}_{ij}$, $\mbox{P}^{S}_{ij}$, $\mbox{P}^{I}_{ij}$ are the probabilities of product $j$ for query $i$ belonging to the exact, substitute, and irrelevant classes, respectively. 
For simplicity, we let $\lambda_1 = \lambda_2 = 0.1$. As a result, the $R_{ij}$ values for the exact match, substitute, and irrelevant classes range from 0.4 to 1, 0.05 to 0.5, and 0 to 0.05, respectively.

\section{Semantic Model}
\label{section.model}
In this section, we present the comprehensive architecture of our proposed EBR, and describe our innovative multi-objective loss function. We also delve into the critical methodologies that contributed significantly to the successful training of our model. These include the introduction of typos, the technique for mining dynamic semi-positives during the negative generation process, and the stratified sampling strategy employed for each batch.

\begin{figure*}
\centering
\includegraphics[width=0.6\textwidth,height=0.35\textheight]{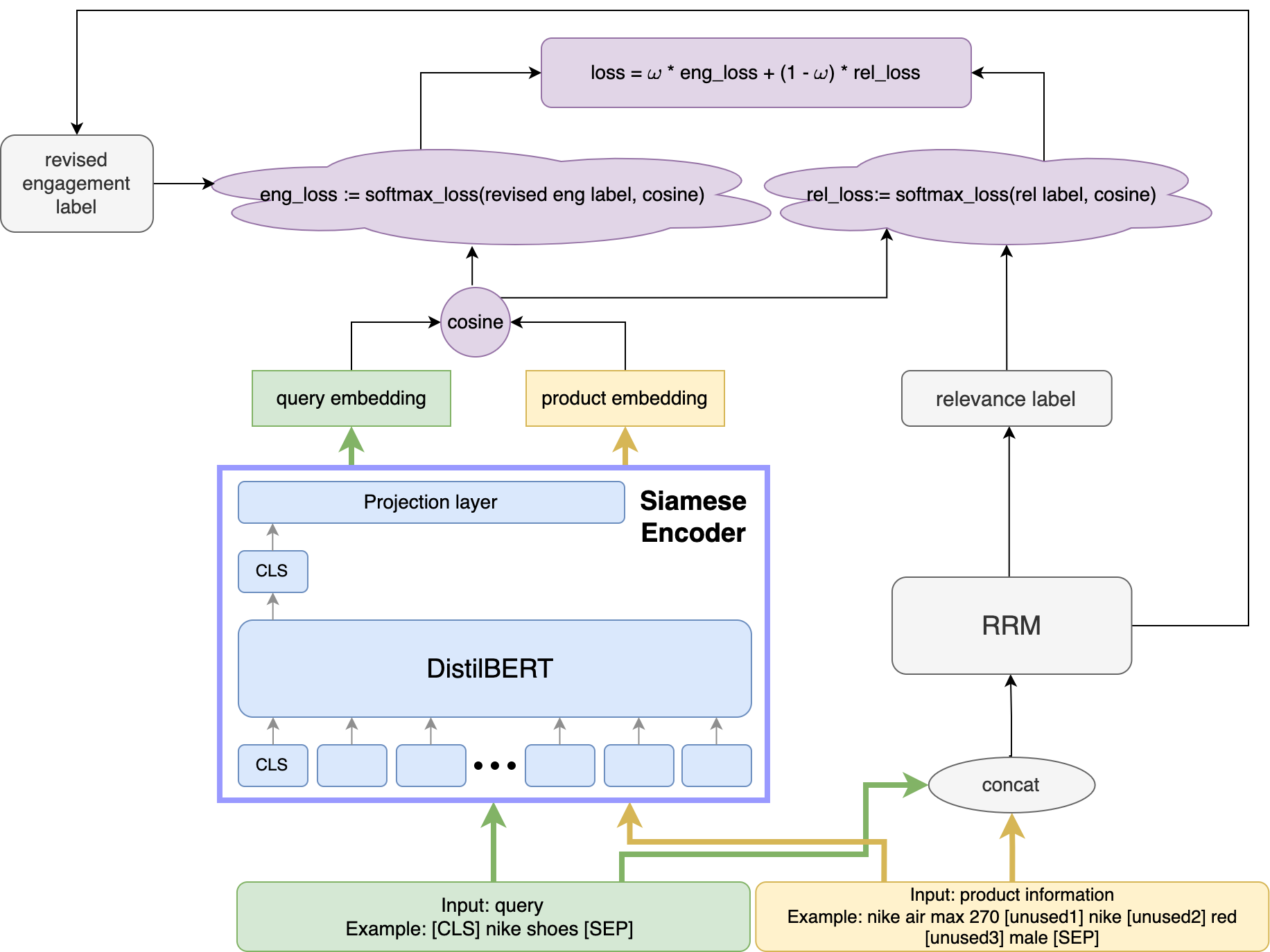}
\caption{Embedding-based retrieval model adopts Siamese dual encoder architecture. Query and document have Blue components are shared between query and products.}
\label{fig.model} 
\end{figure*}

\subsection{Model architecture}
As shown in Figure~\ref{fig.model}, the dual-encoder architecture is adopted for the semantic model, where two separate encoders are used to represent the query and product into the d-dimensional embedding vectors. Then the score of a query and product pair is measured by the cosine similarity of the embeddings. 

The product information consists of a product title, and several attribute values, such as, color, brand, gender, etc. These product attributes are concatenated to the product title using a unique unused token specifically dedicated for each attribute. 


We adopt the Siamese dual-encoder, which has demonstrated superior performance compared to asymmetric dual-encoders \cite{dong2022exploring}. We utilize DistilBERT \cite{sanh2019distilbert} for both the query tower and the product tower. We use their [CLS] token embeddings as representations for queries and products. 

The RRM as described in section~\ref{section.rrm} is employed on the query-product pair to obtain relevance feedback. The RRM inference is conducted for all training instances at once, and the relevance labels are used in the training objectives. 

\subsection{Loss function}
\label{section.loss}
Similar to \cite{magnani2022semantic}, a softmax loss is employed for the products sampled in a stratified way (section~\ref{section.stratified_sampling}) from the positive and negative sets along with their corresponding labels as described in section~\ref{section.label}. The loss is defined as follows
\begin{equation} \label{eq.eng.loss}
lossEng_i = - \sum_j^N \Tilde{S}^\prime_{ij} \log \frac{\exp{\left(\cos{\left(q_i, p_j \right)}/\sigma\right)}}{\sum_j^N \exp{\left(\cos{\left(q_i,p_j\right)}/\sigma \right)}},
\end{equation}
where $N$ is the number of products under consideration, $\Tilde{S}^\prime_{ij} = S^\prime_{ij}/ \sum_j^N S^\prime_{ij}$ denotes the normalized revised label of product $j$ for query $i$, and $q_i$ , $p_j$ are the embedding for query $i$ and product $j$ respectively. $\sigma$ is a temperature factor that is trained together with all the model parameters.

To complement the above engagement-driven loss, we introduced a relevance loss using the relevance label derived from the RRM in section~\ref{section.rel_label}. Similarly, the softmax loss is adopted.
\begin{equation} \label{eq.rel.loss}
lossRel_i = - \sum_j^N \Tilde{R}_{ij} \log \frac{\exp{\left(\cos{\left(q_i, p_j \right)}/\tau\right)}}{\sum_j^N \exp{\left(\cos{\left(q_i,p_j\right)}/\tau \right)}},
\end{equation}
where $\Tilde{R}_{ij} = R_{ij}/\sum_j^N R_{ij}$ represents the normalized relevance label of product $j$ for query $i$ and $q_i$, $\tau$ is a temperature factor that is trained together with all the model parameters.
Finally, the loss for query $i$ is formally expressed as
\begin{equation} \label{eq.loss}
loss_i = \omega * lossEng_i + (1-\omega) * lossRel_i,
\end{equation}
where $\omega $ is an empirically chosen weighting parameter. The selection of $\omega $ is discussed in section~\ref{section.experiments}. 




\subsection{Typos-aware training}
\label{section.typo}
Misspellings frequently occur in e-commerce product searches, with roughly 13\% of user queries containing errors. 
While a spell check model is currently in place on Walmart.com, there are still some instances of misspellings that can be observed. This emphasizes the need for downstream models to be capable of effectively addressing misspellings when they occur.
To address this challenge, we propose a typos-aware training solution in this section.

Similar to \cite{zhuang2021dealing}, we propose incorporating typo injection through query augmentation during the training process to make the model more robust against misspelled queries. In particular, we introduce random space injections within words to emulate the types of queries customers typically submit. For instance, a customer may type "air tag" when searching for "airtag" products. Additionally, we omit typo injection when the token is entirely numerical to prevent queries like "iPhone 14" from being transformed into unrelated product lines. Similar to \cite{zhuang2021dealing}, we utilize the open-source toolkits, TextAttack \cite{morris2020textattack}, to implement the typo generators. The detailed algorithm is presented as follows. 

\begin{itemize}
    \item Perform one noise injection with a 50\% chance
    \item For each noise injection, perform one of the following operations randomly:
    \begin{itemize}
        \item Delete: Delete a random character from the word
        \item Transpose: Swap two adjacent characters in the word
        \item Insert: Insert a random letter in the word
        \item Substitution: Substitute a character in the word with a random letter
        \item Replace: if the selected character has keyboard neighborhoods defined, choose from the adjacent keyboard letters; otherwise, opt from random letters
    \end{itemize}
    \item Skip the typo injection if the injected word is numerical
\end{itemize}


\subsection{Negative sampling schemes}
\label{section.negative}
Selecting appropriate negative samples is crucial for large-scale retrieval tasks, particularly the ability to discern relevant products from hundreds of millions of products in the catalog. Following \cite{magnani2022semantic}, we employ two different sources of negative samples.

\subsubsection{In-batch hard negatives} 
The in-batch negatives are commonly used to generate negative samples for dual-encoders \cite{magnani2022semantic}, as the query and product embeddings are independent and already computed. For a specific query, the product embeddings from other queries in a batch are considered as negative samples. Instead of including all negative samples, we focus on the hardest samples, due to memory constraints, by selecting top-K in-batch products with the highest cosine similarity to the query \cite{liu2021que2search}.  

\subsubsection{Offline hard negatives}
\label{section.offline_negative}
Due to the limited number of queries in a batch, the hard negatives mined in-batch are generally not very informative. To efficiently select negative samples that facilitate model learning, a common approach is ANCE~\cite{xiong2020approximate}, which leverages the retrieval model trained in previous iterations to discover new negatives and construct a different set of examples for the next training iteration. Since negatives are generated from top results using ANN search, false negatives may be included if the model performs well. To find out the relevant products for a given query, \cite{magnani2022semantic} utilizes the product type (PT) attribute of the top products from the training data and the token overlap portion to determine relevance guardrails. The final negatives are then sampled only from products with PTs not present in the relevant PT set and token match rate less than 50\%. However, there are two limitations: First, due to the noisy training data, the top products may not be entirely relevant. Second, the dataset may be too narrow to include all pertinent PTs. To address these limitations, we have implemented a new strategy. 

In our proposed strategy, we employ a product type classifier trained using customer engagement data to predict relevant PTs for each query. We utilize these PT predictions as the relevance guardrails, effectively eliminating false negatives. Moreover, we mine semi-positives from the neighborhood with low positions and reintegrate them into the model training. Lastly, we conduct stratified sampling over PT to ensure a diverse set of negatives. The procedure is described as follows.
\begin{itemize}
\item After completing a training iteration, use the updated model to perform an ANN search on the training dataset.
\item Retrieve the top-K products that are close to query from the ANN search results.
\item Add the product to the \textbf{negative} set with an engagement label of 0, if it satisfies the following conditions.
\begin{itemize}
    \item c1. The product PT is not in the relevant PTs.
    \item c2. The fraction of query tokens found in product title is less than 0.5.
\end{itemize}
\item Add the product to the \textbf{semi-positive} set with an engagement label of $ \min (2, 2*\text{token overlap fraction})$, if it satisfies the following conditions.
\begin{itemize}
    \item c1. The product PT is in the relevant PTs and the PT score is greater than or equal to 0.3.
    \item c2. The fraction of query tokens found in the product tile is greater than or equal to 0.5.
    \item c3. The product has a retrieval position greater than 50.
\end{itemize}
\item Create a new set of training samples by incorporating the newly discovered hard negatives and semi-positives, alongside the positive samples from the previous iterations. 
\end{itemize}

For the negatives and semi-positives collected from the above procedure, engagement labels are assigned as previously mentioned and relevance labels are obtained from RRM. Specifically, negative samples have $S_{ij} = 0$, and $R_{ij}$ values between from 0 to 0.45, similar to the $R_{ij}$s of training set products with $0 < S_{ij} \le 0.01$. For semi-positives, $S_{ij}$ values range from 1 to 2, and $R_{ij}$s from 0.3 to 0.6, comparable to training products with $0.1 < S_{ij} \le 0.8$.



\subsection{Stratified sampling}
\label{section.stratified_sampling}
Due to memory constraints, it is infeasible to incorporate all products for each query into the model training. The batch size is therefore adaptively adjusted to fit within the available GPU memory. In \cite{magnani2022semantic}, products are sampled  randomly for each query regardless of their labels. However, the label distribution per query is usually left-skewed, meaning that the majority of products fall within a low label range. Consequently, if we employ a random sampling strategy, it is very likely that no highly relevant products will be selected. To counteract this, we have implemented a stratified sampling technique for products during training to ensure that the top products for a given query are always relevant. Specifically, for each query, we sample ten products, with four having labels in the range $[1, \infty) $, one with a label in the range $[0.1, 1)$, two with labels in the range $(0, 0.1)$ and the remaining three products being drawn from the negative samples. The stratified sampling is applied on engagement labels.

\section{Experiments and Results}
\label{section.experiments}
\subsection{Dataset and metrics}
\label{section.data}
We gathered engagement data at Walmart.com over a one-year period, from which we sampled $3.7$ million queries with orders. 
For each query, we collected the top $300$ products based on the impressions. In total, our training dataset consists of $780$ million query-product pairs. We divided the dataset into training and validation sets using a $9:1$ ratio. 

To evaluate the relevance, we use three offline evaluation metrics, each applied to one of the three evaluation sets respectively.
We adopt the FAISS library \cite{JDH17} to perform the ANN product search for the offline evaluations. 

\begin{itemize}
    \item \textit{Small index.} We collect $122$k queries from human annotated data. For each query, only the products that were annotated as exact matches are treated as the golden set. The average of recall@K per query, referred to as Exact Match Recall@K, 
    is measured using an ANN retrieval on $3.6$ million products.
    \item \textit{Big index.} We sample $1000$ queries by traffic weight. For each query, we conduct an ANN retrieval on a bigger index products selected at random, approximately $180$ million, and measure the proportion of exact match products, denoted as Exact Match Precision@K, 
    in the top-K results. 
    \item \textit{Purchased dataset.} We collect the $850$k queries with at least one purchase. For each query, only the purchased products are treated as the golden set, regardless their relevance. We calculate the percentage of purchased products retrieved within the top K results, termed Order Recall@K, on a $10$ million product index.
\end{itemize}
We evaluated performance for different retrieval sizes (K = 20, 40, 128, 256, and 512). The results were consistent regardless the choice of K. For brevity, we report the results for K=20. We evaluate the models using different sizes of product index, with \textit{Small index} for rapid iteration and \textit{Big index} to better simulate the online performance. Notably, the evaluation set described in \cite{magnani2022semantic} is similar to the one detailed in the \textit{Purchased dataset}. To measure relevance more accurately, we introduce the first two datasets, in which the metrics are determined solely by the exact match products based on human judgments. To evaluate the relevance performance, we use EM Recall to measure recall and EM Precision to measure precision. For quantifying engagement performance, we utilize Order Recall. We often observe an alignment between EM Recall and EM Precision. However, there is often a trade-off between relevance and engagement metrics.
\subsection{Parameters}
Training is performed on Nvidia A100 GPUs using PyTorch~\cite{NEURIPS2019_9015} and HuggingFace. The models are trained using the Adam~\cite{kingma2014adam} optimizer with a learning rate of $10^{-5}$. The number of products per query is set to $10$, and the number of in-batch negatives is $5$. The batch size is set to $72$. 
We let $\omega$ in loss function (Equation~\ref{eq.loss}) be $0.5$.
\subsection{Offline results}
\label{result.offline}
In this section, we present our offline evaluation results comparing various modeling choices, and detailed ablation studies based on the three metrics described in section~\ref{section.data}. The specific metric is chosen to evaluate different modeling designs. Not all three metrics are presented for each experiment. The results are reported in terms of relative improvement compared to a baseline model.

\begin{figure}
\centering
\includegraphics[width=0.34\textwidth]{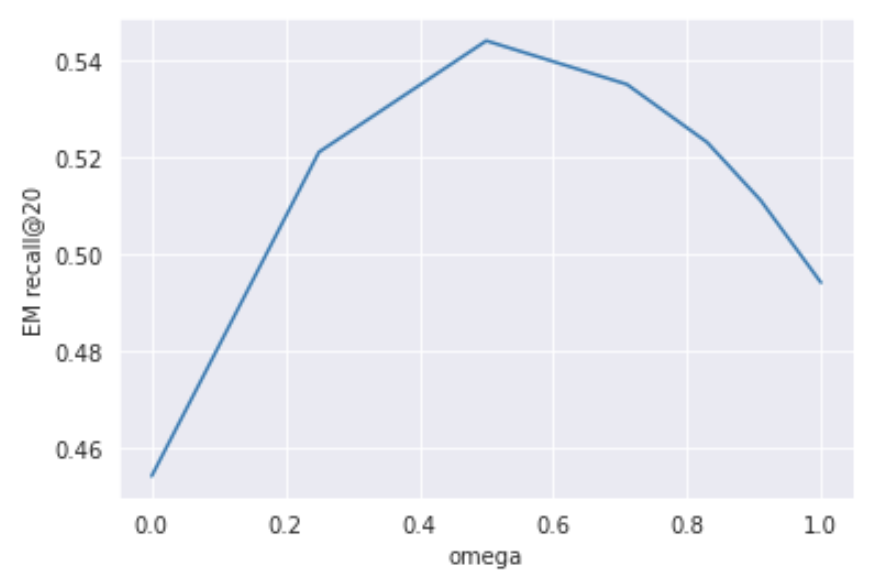}
\caption{EM Recall@20 with respect to various hyper-parameter $\omega$ values}
\label{fig.omega} 
\end{figure}

\begin{table} 
\centering
\small
\begin{tabular}{p{5cm}|p{3cm}} 
 \hline
 Model & Order Recall@20  \\ 
 \hline
  \hline
 DistilBERT proposed in \cite{magnani2022semantic} (baseline) & - \\
  \hline
 baseline + LS & +50.73\%  \\ 
  \hline
 baseline + LS + MOL  & +46.69\% \\
 \hline
\end{tabular}
\caption{Order metrics results for different labeling schemes and losses}
\label{table.scoring}
\end{table}
\begin{table} 
\centering
\small
\begin{tabular}{p{5cm}|p{3cm}} 
 \hline
 Model & EM Recall@20  \\ 
 \hline
  \hline
 DistilBERT proposed in \cite{magnani2022semantic}  (baseline) & - \\
  \hline
 baseline + LR & +2.57\%  \\ 
  \hline
 baseline + MOL  & +7.07\% \\
 \hline
 baseline + LR + MOL  & +9.58\% \\
 \hline
\end{tabular}
\caption{Exact match recall results for various applications of the RRM}
\label{table.rrm}
\end{table}
\begin{table} 
\centering
\small
\begin{tabular}{p{4cm}|p{2.1cm}|p{1.7cm}} 
 \hline
 Model & EM Precision@20 & EM Recall@20 \\ 
 \hline
  \hline
 DistilBERT proposed in \cite{magnani2022semantic} \newline (baseline) & - & - \\ 
 \hline
 baseline + LR & +6.69\% & +2.57\%\\ 
 \hline
 baseline + LR + TI & +7.25\% & +5.78\% \\
 \hline
 baseline + LR + TI + LS + NS & +10.95\% & +12.63\%\\
 \hline
 baseline + LR + TI + LS + NS + MOL & \textbf{+14.65\%} & \textbf{+16.49\%} \\
 \hline
 \end{tabular}
\caption{Results for ablation studies}
\label{table.full_index}
\end{table}

\begin{table*}
\centering
\small
\begin{tabular}{p{1.3cm}|p{4.4cm}|p{1.2cm}|p{2cm}|p{2cm}|p{3cm}} 
 \hline
 Model ID & EBR model description & Control & NDCG@5 lift \newline  (p-value) & NDCG@10 lift \newline  (p-value)  & A/B test revenue lift \newline (p-value)  \\ 
 \hline
  \hline
 I & DistilBERT proposed in \cite{magnani2022semantic} (baseline) & - & - & - & - \\ 
 \hline
 II & baseline + LR & I & \textbf{+1.64\% (0.04)} & \textbf{+1.18\% (0.09)} & - \\ 
 \hline
III & baseline + LR + TI + LS + NS & II & \textbf{+0.64\% (0.10)} & \textbf{+0.68\% (0.09)} & \textbf{+0.31\% (0.01)} \\
 \hline
IV & baseline + LR + TI + LS + NS + MOL & III & +0.50\% (0.12) & \textbf{+0.98\% (0.00)} & \textbf{+0.43\% (0.06)} \\
 \hline
 \end{tabular}
\caption{Online experiment results for the improvements of the EBR model within the hybrid retrieval system}
\label{table.online}
\end{table*}


\subsubsection{Effectiveness of the loss function design}
In our study of the impact of hyper-parameter $\omega$ in Equation~\ref{eq.loss} on recall performance, which balances the engagement loss and relevance loss, we conduct experiments with values ranging from $0$ to $1$, keeping all other parameters constant. The value of $0$ represents optimization based on relevance loss alone, while $1$ focuses on engagement loss. Figure~\ref{fig.omega} reveals that optimizing solely on either the engagement loss or the relevance loss is insufficient. 
The low EM recall metric for small $\omega$s may be due to the RRM's limited performance on deep recall products, especially irrelevant ones. 
The design of both engagement and relevance labels is mutually complementary, with optimal value for $\omega$ of $0.5$. 

\subsubsection{Effectiveness of the new labeling mechanism}
Table~\ref{table.scoring} highlights the influence of the labeling mechanism proposed in Section~\ref{section.label} on the recall metrics of the test set when purchased products serve as the golden set. The baseline is the labeling scheme proposed by \cite{magnani2022semantic}. With all baseline settings and the new labeling scheme (LS), we observe a significant lift of 50.73\% on Order Recall@20. When shifting to the multi-objective loss (MOL), the lift on Order Recall@20 remains noteworthy at around 47\%, albeit slightly lower than the model trained on the engagement loss. This is expected, as an additional relevance label is introduced during the training.



\subsubsection{Effectiveness of two RRM applications}
\label{result.rrm}
We propose utilizing the RRM to the label revisioon (LR) in section~\ref{section.score_augmentation} and the multi-objective loss (MOL) in sections~\ref{section.rel_label}, \ref{section.loss}.
To evaluate the efficacy of these approaches, we conduct experiments with every possible combination and report the metrics in Table~\ref{table.rrm}. The results indicate that both approaches contribute to improving relevance. However, the multi-objective loss has proven to be more effective than label revision, as it impacts every training instances through loss updates. Combining the two approaches retrains the benefits of each, and results in even better performance. 

\subsubsection{Ablation studies}
Table~\ref{table.full_index} presents comprehensive ablation studies detailing the impact of each proposed method, including label revision (LR), typo injection (TI), new labeling scheme (LS), new negative sampling (NS), and multi-objective loss (MOL). 
With fewer false positives within training data by implementing label revision (section~\ref{section.score_augmentation}), we surpass the baseline by 6.69\% in terms of the percentage of exact match products within the top 20 retrieved items. Additional type-aware training (section~\ref{section.typo}) for handling misspelled queries, provides an extra boost of 0.5\%. The introduction of a new labeling scheme (section~\ref{section.label}) and an enhanced negative sampling algorithm (section~\ref{section.negative}) for reducing training false negatives further elevates EM Precision@20 by an additional 3.7\%. Finally, by training with the proposed multi-objective loss (section~\ref{section.loss}), we achieve an extra lift of 3.7\%. Similar patterns are observed in the EM Recall@20 metrics. The most significant improvement comes from the use of multi-objective loss, though each enhancement also exhibits an increase in relevance metrics.

\subsection{Online Experiments} 
\label{result.online}

We deploy our models online to evaluate the performance of our proposed methodologies. As in \cite{magnani2022semantic}, we utilize a hybrid retrieval system that retrieves products from both EBR and traditional lexical retrieval. Throughout the experiments, we maintain the lexical retrieval component unchanged, focusing solely on examining the improvements of the EBR model. 
The evaluation is done by assessing the relevance for the products ranked in top 10 end-to-end for impacted query and conducting A/B testing on live Walmart traffic. 
For the relevance evaluation, human judgments are scaled based on three classes (exact, substitute, irrelevant), derived from the product page, title, and image. The NDCG metrics are computed based on this three-point scale. For the A/B tests, the revenue lifts are reported. The results can be found in Table~\ref{table.online}. Model II was not subjected to A/B testing due to its small traffic impact. 
It's crucial to note that the metrics for these models are not directly comparable to those in previous tables. The earlier metrics represent relative improvements to the baseline, whereas in our online test, each model was released sequentially, meaning that the reported metrics reflect increments from the preceding model. 
Nearly all models exhibit statistically positive results ($\alpha=0.1$) in NDCGs, and revenue. The positive results of our online test further validate the effectiveness of the proposed new techniques. In particular, these results indicate that even when a stronger baseline model, which includes a lexical component in addition to EBR, is employed, the new techniques consistently demonstrate effectiveness, suggesting their robustness.

\section{Conclusion}
In this paper, we introduce several techniques that enhance the relevance of a state-of-the-art EBR retrieval model deployed on Walmart.com by efficiently addressing existing challenges, such as data issues and query typos.
Among the proposed methods, the innovative multi-objective loss has shown the most relevance enhancement. 
Those new EBR models have proven to be superior to the state-of-the-art EBR \cite{magnani2022semantic}, and have been successfully launched to serve live traffic on Walmart.com.

\section*{Acknowledgments}
The authors would like to express their gratitude to Professor ChengXiang Zhai for his valuable insights and his generous support during our discussions.
\newpage
\vfill\eject 
\bibliographystyle{ACM-Reference-Format}
\balance
\bibliography{references}


\begin{thebibliography}{34}


\ifx \showCODEN    \undefined \def \showCODEN     #1{\unskip}     \fi
\ifx \showDOI      \undefined \def \showDOI       #1{#1}\fi
\ifx \showISBNx    \undefined \def \showISBNx     #1{\unskip}     \fi
\ifx \showISBNxiii \undefined \def \showISBNxiii  #1{\unskip}     \fi
\ifx \showISSN     \undefined \def \showISSN      #1{\unskip}     \fi
\ifx \showLCCN     \undefined \def \showLCCN      #1{\unskip}     \fi
\ifx \shownote     \undefined \def \shownote      #1{#1}          \fi
\ifx \showarticletitle \undefined \def \showarticletitle #1{#1}   \fi
\ifx \showURL      \undefined \def \showURL       {\relax}        \fi
\providecommand\bibfield[2]{#2}
\providecommand\bibinfo[2]{#2}
\providecommand\natexlab[1]{#1}
\providecommand\showeprint[2][]{arXiv:#2}

\bibitem[Bromley et~al\mbox{.}(1993)]%
        {bromley1993signature}
\bibfield{author}{\bibinfo{person}{Jane Bromley}, \bibinfo{person}{James~W Bentz}, \bibinfo{person}{L{\'e}on Bottou}, \bibinfo{person}{Isabelle Guyon}, \bibinfo{person}{Yann LeCun}, \bibinfo{person}{Cliff Moore}, \bibinfo{person}{Eduard S{\"a}ckinger}, {and} \bibinfo{person}{Roopak Shah}.} \bibinfo{year}{1993}\natexlab{}.
\newblock \showarticletitle{Signature verification using a “siamese” time delay neural network}.
\newblock \bibinfo{journal}{\emph{IJPRAI}} \bibinfo{volume}{7}, \bibinfo{number}{04} (\bibinfo{year}{1993}), \bibinfo{pages}{669--688}.
\newblock


\bibitem[Cao et~al\mbox{.}(2007)]%
        {cao2007learning}
\bibfield{author}{\bibinfo{person}{Zhe Cao}, \bibinfo{person}{Tao Qin}, \bibinfo{person}{Tie-Yan Liu}, \bibinfo{person}{Ming-Feng Tsai}, {and} \bibinfo{person}{Hang Li}.} \bibinfo{year}{2007}\natexlab{}.
\newblock \showarticletitle{Learning to rank: from pairwise approach to listwise approach}. In \bibinfo{booktitle}{\emph{Proceedings of the 24th international conference on Machine learning}}. \bibinfo{pages}{129--136}.
\newblock


\bibitem[Dai and Callan(2019)]%
        {dai2019deeper}
\bibfield{author}{\bibinfo{person}{Zhuyun Dai} {and} \bibinfo{person}{Jamie Callan}.} \bibinfo{year}{2019}\natexlab{}.
\newblock \showarticletitle{Deeper text understanding for IR with contextual neural language modeling}. In \bibinfo{booktitle}{\emph{Proceedings of the 42nd international ACM SIGIR conference on research and development in information retrieval}}. \bibinfo{pages}{985--988}.
\newblock


\bibitem[Dong et~al\mbox{.}(2022)]%
        {dong2022exploring}
\bibfield{author}{\bibinfo{person}{Zhe Dong}, \bibinfo{person}{Jianmo Ni}, \bibinfo{person}{Daniel~M Bikel}, \bibinfo{person}{Enrique Alfonseca}, \bibinfo{person}{Yuan Wang}, \bibinfo{person}{Chen Qu}, {and} \bibinfo{person}{Imed Zitouni}.} \bibinfo{year}{2022}\natexlab{}.
\newblock \showarticletitle{Exploring dual encoder architectures for question answering}.
\newblock \bibinfo{journal}{\emph{arXiv preprint arXiv:2204.07120}} (\bibinfo{year}{2022}).
\newblock


\bibitem[Gao et~al\mbox{.}(2021)]%
        {gao2021complement}
\bibfield{author}{\bibinfo{person}{Luyu Gao}, \bibinfo{person}{Zhuyun Dai}, \bibinfo{person}{Tongfei Chen}, \bibinfo{person}{Zhen Fan}, \bibinfo{person}{Benjamin~Van Durme}, {and} \bibinfo{person}{Jamie Callan}.} \bibinfo{year}{2021}\natexlab{}.
\newblock \showarticletitle{Complement lexical retrieval model with semantic residual embeddings}. In \bibinfo{booktitle}{\emph{European Conference on Information Retrieval}}. Springer, \bibinfo{pages}{146--160}.
\newblock


\bibitem[He et~al\mbox{.}(2023)]%
        {he2023que2engage}
\bibfield{author}{\bibinfo{person}{Yunzhong He}, \bibinfo{person}{Yuxin Tian}, \bibinfo{person}{Mengjiao Wang}, \bibinfo{person}{Feier Chen}, \bibinfo{person}{Licheng Yu}, \bibinfo{person}{Maolong Tang}, \bibinfo{person}{Congcong Chen}, \bibinfo{person}{Ning Zhang}, \bibinfo{person}{Bin Kuang}, {and} \bibinfo{person}{Arul Prakash}.} \bibinfo{year}{2023}\natexlab{}.
\newblock \showarticletitle{Que2Engage: Embedding-based Retrieval for Relevant and Engaging Products at Facebook Marketplace}.
\newblock \bibinfo{journal}{\emph{arXiv preprint arXiv:2302.11052}} (\bibinfo{year}{2023}).
\newblock


\bibitem[Huang et~al\mbox{.}(2013)]%
        {DSSM}
\bibfield{author}{\bibinfo{person}{Po{-}Sen Huang}, \bibinfo{person}{Xiaodong He}, \bibinfo{person}{Jianfeng Gao}, \bibinfo{person}{Li Deng}, \bibinfo{person}{Alex Acero}, {and} \bibinfo{person}{Larry~P. Heck}.} \bibinfo{year}{2013}\natexlab{}.
\newblock \showarticletitle{Learning deep structured semantic models for web search using clickthrough data}. In \bibinfo{booktitle}{\emph{CIKM, 2013}}. \bibinfo{pages}{2333--2338}.
\newblock


\bibitem[Jha et~al\mbox{.}(2023)]%
        {jha2023unified}
\bibfield{author}{\bibinfo{person}{Rishikesh Jha}, \bibinfo{person}{Siddharth Subramaniyam}, \bibinfo{person}{Ethan Benjamin}, {and} \bibinfo{person}{Thrivikrama Taula}.} \bibinfo{year}{2023}\natexlab{}.
\newblock \showarticletitle{Unified Embedding Based Personalized Retrieval in Etsy Search}.
\newblock \bibinfo{journal}{\emph{arXiv preprint arXiv:2306.04833}} (\bibinfo{year}{2023}).
\newblock


\bibitem[Johnson et~al\mbox{.}(2017)]%
        {JDH17}
\bibfield{author}{\bibinfo{person}{Jeff Johnson}, \bibinfo{person}{Matthijs Douze}, {and} \bibinfo{person}{Herv{\'e} J{\'e}gou}.} \bibinfo{year}{2017}\natexlab{}.
\newblock \showarticletitle{Billion-scale similarity search with GPUs}.
\newblock \bibinfo{journal}{\emph{arXiv preprint arXiv:1702.08734}} (\bibinfo{year}{2017}).
\newblock


\bibitem[Johnson et~al\mbox{.}(2019)]%
        {johnson2019billion}
\bibfield{author}{\bibinfo{person}{Jeff Johnson}, \bibinfo{person}{Matthijs Douze}, {and} \bibinfo{person}{Herv{\'e} J{\'e}gou}.} \bibinfo{year}{2019}\natexlab{}.
\newblock \showarticletitle{Billion-scale similarity search with gpus}.
\newblock \bibinfo{journal}{\emph{IEEE Transactions on Big Data}} \bibinfo{volume}{7}, \bibinfo{number}{3} (\bibinfo{year}{2019}), \bibinfo{pages}{535--547}.
\newblock


\bibitem[Kingma and Ba(2014)]%
        {kingma2014adam}
\bibfield{author}{\bibinfo{person}{Diederik~P Kingma} {and} \bibinfo{person}{Jimmy Ba}.} \bibinfo{year}{2014}\natexlab{}.
\newblock \showarticletitle{Adam: A method for stochastic optimization}.
\newblock \bibinfo{journal}{\emph{arXiv preprint arXiv:1412.6980}} (\bibinfo{year}{2014}).
\newblock


\bibitem[Li et~al\mbox{.}(2021)]%
        {li2021embedding}
\bibfield{author}{\bibinfo{person}{Sen Li}, \bibinfo{person}{Fuyu Lv}, \bibinfo{person}{Taiwei Jin}, \bibinfo{person}{Guli Lin}, \bibinfo{person}{Keping Yang}, \bibinfo{person}{Xiaoyi Zeng}, \bibinfo{person}{Xiao-Ming Wu}, {and} \bibinfo{person}{Qianli Ma}.} \bibinfo{year}{2021}\natexlab{}.
\newblock \showarticletitle{Embedding-based product retrieval in taobao search}. In \bibinfo{booktitle}{\emph{Proceedings of the 27th ACM SIGKDD Conference on Knowledge Discovery \& Data Mining}}. \bibinfo{pages}{3181--3189}.
\newblock


\bibitem[Liu et~al\mbox{.}(2021)]%
        {liu2021que2search}
\bibfield{author}{\bibinfo{person}{Yiqun Liu}, \bibinfo{person}{Kaushik Rangadurai}, \bibinfo{person}{Yunzhong He}, \bibinfo{person}{Siddarth Malreddy}, \bibinfo{person}{Xunlong Gui}, \bibinfo{person}{Xiaoyi Liu}, {and} \bibinfo{person}{Fedor Borisyuk}.} \bibinfo{year}{2021}\natexlab{}.
\newblock \showarticletitle{Que2search: Fast and accurate query and document understanding for search at facebook}. In \bibinfo{booktitle}{\emph{Proceedings of the 27th ACM SIGKDD Conference on Knowledge Discovery \& Data Mining}}. \bibinfo{pages}{3376--3384}.
\newblock


\bibitem[Lu et~al\mbox{.}(2022)]%
        {lu2022ernie}
\bibfield{author}{\bibinfo{person}{Yuxiang Lu}, \bibinfo{person}{Yiding Liu}, \bibinfo{person}{Jiaxiang Liu}, \bibinfo{person}{Yunsheng Shi}, \bibinfo{person}{Zhengjie Huang}, \bibinfo{person}{Shikun Feng~Yu Sun}, \bibinfo{person}{Hao Tian}, \bibinfo{person}{Hua Wu}, \bibinfo{person}{Shuaiqiang Wang}, \bibinfo{person}{Dawei Yin}, {et~al\mbox{.}}} \bibinfo{year}{2022}\natexlab{}.
\newblock \showarticletitle{Ernie-search: Bridging cross-encoder with dual-encoder via self on-the-fly distillation for dense passage retrieval}.
\newblock \bibinfo{journal}{\emph{arXiv preprint arXiv:2205.09153}} (\bibinfo{year}{2022}).
\newblock


\bibitem[Magnani et~al\mbox{.}(2022)]%
        {magnani2022semantic}
\bibfield{author}{\bibinfo{person}{Alessandro Magnani}, \bibinfo{person}{Feng Liu}, \bibinfo{person}{Suthee Chaidaroon}, \bibinfo{person}{Sachin Yadav}, \bibinfo{person}{Praveen Reddy~Suram}, \bibinfo{person}{Ajit Puthenputhussery}, \bibinfo{person}{Sijie Chen}, \bibinfo{person}{Min Xie}, \bibinfo{person}{Anirudh Kashi}, \bibinfo{person}{Tony Lee}, {et~al\mbox{.}}} \bibinfo{year}{2022}\natexlab{}.
\newblock \showarticletitle{Semantic retrieval at walmart}. In \bibinfo{booktitle}{\emph{Proceedings of the 28th ACM SIGKDD Conference on Knowledge Discovery and Data Mining}}. \bibinfo{pages}{3495--3503}.
\newblock


\bibitem[Manning et~al\mbox{.}(2008)]%
        {ir}
\bibfield{author}{\bibinfo{person}{Christopher~D. Manning}, \bibinfo{person}{Prabhakar Raghavan}, {and} \bibinfo{person}{Hinrich Sch{\"{u}}tze}.} \bibinfo{year}{2008}\natexlab{}.
\newblock \bibinfo{booktitle}{\emph{Introduction to information retrieval}}.
\newblock \bibinfo{publisher}{Cambridge University Press}.
\newblock
\showISBNx{978-0-521-86571-5}


\bibitem[Morris et~al\mbox{.}(2020)]%
        {morris2020textattack}
\bibfield{author}{\bibinfo{person}{John~X Morris}, \bibinfo{person}{Eli Lifland}, \bibinfo{person}{Jin~Yong Yoo}, \bibinfo{person}{Jake Grigsby}, \bibinfo{person}{Di Jin}, {and} \bibinfo{person}{Yanjun Qi}.} \bibinfo{year}{2020}\natexlab{}.
\newblock \showarticletitle{Textattack: A framework for adversarial attacks, data augmentation, and adversarial training in nlp}.
\newblock \bibinfo{journal}{\emph{arXiv preprint arXiv:2005.05909}} (\bibinfo{year}{2020}).
\newblock


\bibitem[Muhamed et~al\mbox{.}(2023)]%
        {muhamed2023web}
\bibfield{author}{\bibinfo{person}{Aashiq Muhamed}, \bibinfo{person}{Sriram Srinivasan}, \bibinfo{person}{Choon-Hui Teo}, \bibinfo{person}{Qingjun Cui}, \bibinfo{person}{Belinda Zeng}, \bibinfo{person}{Trishul Chilimbi}, {and} \bibinfo{person}{SVN Vishwanathan}.} \bibinfo{year}{2023}\natexlab{}.
\newblock \showarticletitle{Web-scale semantic product search with large language models}. In \bibinfo{booktitle}{\emph{Pacific-Asia Conference on Knowledge Discovery and Data Mining}}. Springer, \bibinfo{pages}{73--85}.
\newblock


\bibitem[Nigam et~al\mbox{.}(2019)]%
        {nigam2019semantic}
\bibfield{author}{\bibinfo{person}{Priyanka Nigam}, \bibinfo{person}{Yiwei Song}, \bibinfo{person}{Vijai Mohan}, \bibinfo{person}{Vihan Lakshman}, \bibinfo{person}{Weitian Ding}, \bibinfo{person}{Ankit Shingavi}, \bibinfo{person}{Choon~Hui Teo}, \bibinfo{person}{Hao Gu}, {and} \bibinfo{person}{Bing Yin}.} \bibinfo{year}{2019}\natexlab{}.
\newblock \showarticletitle{Semantic product search}. In \bibinfo{booktitle}{\emph{Proceedings of the 25th ACM SIGKDD International Conference on Knowledge Discovery \& Data Mining}}. \bibinfo{pages}{2876--2885}.
\newblock


\bibitem[Nogueira and Cho(2019)]%
        {nogueira2019passage}
\bibfield{author}{\bibinfo{person}{Rodrigo Nogueira} {and} \bibinfo{person}{Kyunghyun Cho}.} \bibinfo{year}{2019}\natexlab{}.
\newblock \showarticletitle{Passage Re-ranking with BERT}.
\newblock \bibinfo{journal}{\emph{arXiv preprint arXiv:1901.04085}} (\bibinfo{year}{2019}).
\newblock


\bibitem[Ouyang et~al\mbox{.}(2022)]%
        {ouyang2022training}
\bibfield{author}{\bibinfo{person}{Long Ouyang}, \bibinfo{person}{Jeffrey Wu}, \bibinfo{person}{Xu Jiang}, \bibinfo{person}{Diogo Almeida}, \bibinfo{person}{Carroll Wainwright}, \bibinfo{person}{Pamela Mishkin}, \bibinfo{person}{Chong Zhang}, \bibinfo{person}{Sandhini Agarwal}, \bibinfo{person}{Katarina Slama}, \bibinfo{person}{Alex Ray}, {et~al\mbox{.}}} \bibinfo{year}{2022}\natexlab{}.
\newblock \showarticletitle{Training language models to follow instructions with human feedback}.
\newblock \bibinfo{journal}{\emph{Advances in Neural Information Processing Systems}}  \bibinfo{volume}{35} (\bibinfo{year}{2022}), \bibinfo{pages}{27730--27744}.
\newblock


\bibitem[Paszke et~al\mbox{.}(2019)]%
        {NEURIPS2019_9015}
\bibfield{author}{\bibinfo{person}{Adam Paszke}, \bibinfo{person}{Sam Gross}, \bibinfo{person}{Francisco Massa}, \bibinfo{person}{Adam Lerer}, \bibinfo{person}{James Bradbury}, \bibinfo{person}{Gregory Chanan}, \bibinfo{person}{Trevor Killeen}, \bibinfo{person}{Zeming Lin}, \bibinfo{person}{Natalia Gimelshein}, \bibinfo{person}{Luca Antiga}, \bibinfo{person}{Alban Desmaison}, \bibinfo{person}{Andreas Kopf}, \bibinfo{person}{Edward Yang}, \bibinfo{person}{Zachary DeVito}, \bibinfo{person}{Martin Raison}, \bibinfo{person}{Alykhan Tejani}, \bibinfo{person}{Sasank Chilamkurthy}, \bibinfo{person}{Benoit Steiner}, \bibinfo{person}{Lu Fang}, \bibinfo{person}{Junjie Bai}, {and} \bibinfo{person}{Soumith Chintala}.} \bibinfo{year}{2019}\natexlab{}.
\newblock \showarticletitle{PyTorch: An Imperative Style, High-Performance Deep Learning Library}.
\newblock In \bibinfo{booktitle}{\emph{Advances in Neural Information Processing Systems 32}}, \bibfield{editor}{\bibinfo{person}{H.~Wallach}, \bibinfo{person}{H.~Larochelle}, \bibinfo{person}{A.~Beygelzimer}, \bibinfo{person}{F.~d\textquotesingle Alch\'{e}-Buc}, \bibinfo{person}{E.~Fox}, {and} \bibinfo{person}{R.~Garnett}} (Eds.). \bibinfo{publisher}{Curran Associates, Inc.}, \bibinfo{pages}{8024--8035}.
\newblock
\urldef\tempurl%
\url{http://papers.neurips.cc/paper/9015-pytorch-an-imperative-style-high-performance-deep-learning-library.pdf}
\showURL{%
\tempurl}


\bibitem[Qu et~al\mbox{.}(2020)]%
        {qu2020rocketqa}
\bibfield{author}{\bibinfo{person}{Yingqi Qu}, \bibinfo{person}{Yuchen Ding}, \bibinfo{person}{Jing Liu}, \bibinfo{person}{Kai Liu}, \bibinfo{person}{Ruiyang Ren}, \bibinfo{person}{Wayne~Xin Zhao}, \bibinfo{person}{Daxiang Dong}, \bibinfo{person}{Hua Wu}, {and} \bibinfo{person}{Haifeng Wang}.} \bibinfo{year}{2020}\natexlab{}.
\newblock \showarticletitle{RocketQA: An optimized training approach to dense passage retrieval for open-domain question answering}.
\newblock \bibinfo{journal}{\emph{arXiv preprint arXiv:2010.08191}} (\bibinfo{year}{2020}).
\newblock


\bibitem[Reddy et~al\mbox{.}(2022)]%
        {reddy2022shopping}
\bibfield{author}{\bibinfo{person}{Chandan~K. Reddy}, \bibinfo{person}{Lluís Màrquez}, \bibinfo{person}{Fran Valero}, \bibinfo{person}{Nikhil Rao}, \bibinfo{person}{Hugo Zaragoza}, \bibinfo{person}{Sambaran Bandyopadhyay}, \bibinfo{person}{Arnab Biswas}, \bibinfo{person}{Anlu Xing}, {and} \bibinfo{person}{Karthik Subbian}.} \bibinfo{year}{2022}\natexlab{}.
\newblock \showarticletitle{Shopping Queries Dataset: A Large-Scale {ESCI} Benchmark for Improving Product Search}.
\newblock  (\bibinfo{year}{2022}).
\newblock
\showeprint[arxiv]{2206.06588}


\bibitem[Rossi et~al\mbox{.}(2024)]%
        {rossi2024relevance}
\bibfield{author}{\bibinfo{person}{Nicholas Rossi}, \bibinfo{person}{Juexin Lin}, \bibinfo{person}{Feng Liu}, \bibinfo{person}{Zhen Yang}, \bibinfo{person}{Tony Lee}, \bibinfo{person}{Alessandro Magnani}, {and} \bibinfo{person}{Ciya Liao}.} \bibinfo{year}{2024}\natexlab{}.
\newblock \showarticletitle{Relevance Filtering for Embedding-based Retrieval}. In \bibinfo{booktitle}{\emph{Proceedings of the 33rd ACM International Conference on Information and Knowledge Management (CIKM '24)}}.
\newblock
\urldef\tempurl%
\url{https://doi.org/10.1145/3627673.3680095}
\showDOI{\tempurl}


\bibitem[Sanh et~al\mbox{.}(2019)]%
        {sanh2019distilbert}
\bibfield{author}{\bibinfo{person}{Victor Sanh}, \bibinfo{person}{Lysandre Debut}, \bibinfo{person}{Julien Chaumond}, {and} \bibinfo{person}{Thomas Wolf}.} \bibinfo{year}{2019}\natexlab{}.
\newblock \showarticletitle{DistilBERT, a distilled version of BERT: smaller, faster, cheaper and lighter}.
\newblock \bibinfo{journal}{\emph{arXiv preprint arXiv:1910.01108}} (\bibinfo{year}{2019}).
\newblock


\bibitem[Xiong et~al\mbox{.}(2020)]%
        {xiong2020approximate}
\bibfield{author}{\bibinfo{person}{Lee Xiong}, \bibinfo{person}{Chenyan Xiong}, \bibinfo{person}{Ye Li}, \bibinfo{person}{Kwok-Fung Tang}, \bibinfo{person}{Jialin Liu}, \bibinfo{person}{Paul Bennett}, \bibinfo{person}{Junaid Ahmed}, {and} \bibinfo{person}{Arnold Overwijk}.} \bibinfo{year}{2020}\natexlab{}.
\newblock \showarticletitle{Approximate nearest neighbor negative contrastive learning for dense text retrieval}.
\newblock \bibinfo{journal}{\emph{arXiv preprint arXiv:2007.00808}} (\bibinfo{year}{2020}).
\newblock


\bibitem[Yang et~al\mbox{.}(2019)]%
        {yang2019simple}
\bibfield{author}{\bibinfo{person}{Wei Yang}, \bibinfo{person}{Haotian Zhang}, {and} \bibinfo{person}{Jimmy Lin}.} \bibinfo{year}{2019}\natexlab{}.
\newblock \showarticletitle{Simple applications of BERT for ad hoc document retrieval}.
\newblock \bibinfo{journal}{\emph{arXiv preprint arXiv:1903.10972}} (\bibinfo{year}{2019}).
\newblock


\bibitem[Yao et~al\mbox{.}(2021)]%
        {yao2021learning}
\bibfield{author}{\bibinfo{person}{Shaowei Yao}, \bibinfo{person}{Jiwei Tan}, \bibinfo{person}{Xi Chen}, \bibinfo{person}{Keping Yang}, \bibinfo{person}{Rong Xiao}, \bibinfo{person}{Hongbo Deng}, {and} \bibinfo{person}{Xiaojun Wan}.} \bibinfo{year}{2021}\natexlab{}.
\newblock \showarticletitle{Learning a product relevance model from click-through data in e-commerce}. In \bibinfo{booktitle}{\emph{Proceedings of the Web Conference 2021}}. \bibinfo{pages}{2890--2899}.
\newblock


\bibitem[Zhan et~al\mbox{.}(2021)]%
        {zhan2021optimizing}
\bibfield{author}{\bibinfo{person}{Jingtao Zhan}, \bibinfo{person}{Jiaxin Mao}, \bibinfo{person}{Yiqun Liu}, \bibinfo{person}{Jiafeng Guo}, \bibinfo{person}{Min Zhang}, {and} \bibinfo{person}{Shaoping Ma}.} \bibinfo{year}{2021}\natexlab{}.
\newblock \showarticletitle{Optimizing dense retrieval model training with hard negatives}. In \bibinfo{booktitle}{\emph{Proceedings of the 44th International ACM SIGIR Conference on Research and Development in Information Retrieval}}. \bibinfo{pages}{1503--1512}.
\newblock


\bibitem[Zhang et~al\mbox{.}(2020)]%
        {zhang2020towards}
\bibfield{author}{\bibinfo{person}{Han Zhang}, \bibinfo{person}{Songlin Wang}, \bibinfo{person}{Kang Zhang}, \bibinfo{person}{Zhiling Tang}, \bibinfo{person}{Yunjiang Jiang}, \bibinfo{person}{Yun Xiao}, \bibinfo{person}{Weipeng Yan}, {and} \bibinfo{person}{Wen-Yun Yang}.} \bibinfo{year}{2020}\natexlab{}.
\newblock \showarticletitle{Towards personalized and semantic retrieval: An end-to-end solution for e-commerce search via embedding learning}. In \bibinfo{booktitle}{\emph{Proceedings of the 43rd International ACM SIGIR Conference on Research and Development in Information Retrieval}}. \bibinfo{pages}{2407--2416}.
\newblock


\bibitem[Zheng et~al\mbox{.}(2023)]%
        {zheng2023delving}
\bibfield{author}{\bibinfo{person}{Xiaoyang Zheng}, \bibinfo{person}{Fuyu Lv}, \bibinfo{person}{Zilong Wang}, \bibinfo{person}{Qingwen Liu}, {and} \bibinfo{person}{Xiaoyi Zeng}.} \bibinfo{year}{2023}\natexlab{}.
\newblock \showarticletitle{Delving into E-Commerce Product Retrieval with Vision-Language Pre-training}.
\newblock \bibinfo{journal}{\emph{arXiv preprint arXiv:2304.04377}} (\bibinfo{year}{2023}).
\newblock


\bibitem[Zheng et~al\mbox{.}(2022)]%
        {zheng2022multi}
\bibfield{author}{\bibinfo{person}{Yukun Zheng}, \bibinfo{person}{Jiang Bian}, \bibinfo{person}{Guanghao Meng}, \bibinfo{person}{Chao Zhang}, \bibinfo{person}{Honggang Wang}, \bibinfo{person}{Zhixuan Zhang}, \bibinfo{person}{Sen Li}, \bibinfo{person}{Tao Zhuang}, \bibinfo{person}{Qingwen Liu}, {and} \bibinfo{person}{Xiaoyi Zeng}.} \bibinfo{year}{2022}\natexlab{}.
\newblock \showarticletitle{Multi-Objective Personalized Product Retrieval in Taobao Search}.
\newblock \bibinfo{journal}{\emph{arXiv preprint arXiv:2210.04170}} (\bibinfo{year}{2022}).
\newblock


\bibitem[Zhuang and Zuccon(2021)]%
        {zhuang2021dealing}
\bibfield{author}{\bibinfo{person}{Shengyao Zhuang} {and} \bibinfo{person}{Guido Zuccon}.} \bibinfo{year}{2021}\natexlab{}.
\newblock \showarticletitle{Dealing with typos for BERT-based passage retrieval and ranking}.
\newblock \bibinfo{journal}{\emph{arXiv preprint arXiv:2108.12139}} (\bibinfo{year}{2021}).
\newblock


\end{thebibliography}

\end{document}